\documentclass[prl,twocolumn,lettersize,preprintnumbers,amsmath,amssymb]{revtex4}
\usepackage{graphicx}
\usepackage{latexsym}
\usepackage{amsmath}
\usepackage{graphics}
\usepackage{amssymb}
\usepackage{layout}
\usepackage{verbatim}
\usepackage{amsfonts}
\usepackage{epsfig}
\usepackage{slashed}
\usepackage{feynmp}
\usepackage{float}
\usepackage{epstopdf}
\newcommand{\ket}[1]{\left|#1\right>}

\newcommand{\beq}{\begin{equation}}
\newcommand{\eeq}{\end{equation}}
\newcommand{\bea}{\begin{eqnarray}}
\newcommand{\eea}{\end{eqnarray}}
\newcommand{\nn}{\nonumber}

\begin{document}

\title{Filter function formalism beyond pure dephasing and non-Markovian noise in singlet-triplet qubits}
\author{Edwin Barnes,$^{1,2}$ Mark S. Rudner,$^{3}$ Frederico Martins,$^{3}$ Filip K. Malinowski,$^{3}$ Charles M. Marcus,$^{3}$ and Ferdinand Kuemmeth$^{3}$}
\affiliation{$^{1}$Department of Physics, Virginia Tech, Blacksburg, Virginia 24061, USA\\
$^{2}$Condensed Matter Theory Center and Joint Quantum Institute, Department of Physics, University of Maryland, College Park, Maryland 20742-4111, USA\\
$^{3}$Center for Quantum Devices, Niels Bohr Institute, University of Copenhagen, 2100 Copenhagen, Denmark}

\begin{abstract}
The filter function formalism quantitatively describes the dephasing of a qubit by a bath that causes Gaussian fluctuations in the qubit energies with an arbitrary noise power spectrum. Here, we extend this formalism to account for more general types of noise that couple to the qubit through terms that do not commute with the qubit's bare Hamiltonian. Our approach applies to any power spectrum that generates slow noise fluctuations in the qubit's evolution. We demonstrate our formalism in the case of singlet-triplet qubits subject to both quasistatic nuclear noise and $1/\omega^\alpha$ charge noise and find good agreement with recent experimental findings. This comparison shows the efficacy of our approach in describing real systems and additionally highlights the challenges with distinguishing different types of noise in free induction decay experiments.
\end{abstract}

\maketitle

Decoherence presents an important challenge for quantum-based technologies and is particularly relevant for solid state nanoscale devices. A crucial step in overcoming this challenge is to understand how various noise sources affect a qubit's evolution. It is often the case that noise caused by e.g., nuclear spin or charge fluctuations in quantum dots, or flux noise in superconducting circuits, can be well described by a classical Gaussian ensemble characterized by an appropriate power spectrum \cite{Medford_PRL12,Dial_PRL13,Bialczak_PRL07}. In the case of pure dephasing where the noise creates fluctuations in the qubit energy levels, considerable progress has been made in understanding how the qubit responds to a given noise spectrum \cite{Ithier_PRB05,Cywinski_PRB08,Green_PRL12,Cywinski_PRA14}. However, it is often the case that noise induces not only energy fluctuations but also rotations between the qubit states. A quantitative theory of how the qubit evolves in the presence of this more general type of decoherence could be used to better understand noise sources and develop new approaches to dynamical decoupling \cite{Uhrig_PRL07,Witzel_PRL07,Bluhm_NP11,Malinowski_arxiv16} and dynamically corrected gates \cite{Wang_NatComm12,Kestner_PRL13,Wang_PRB14,Barnes_SciRep15}.

In this work, we take an important step toward addressing this problem by extending the filter function approach to treat more general types of decoherence beyond the case of pure dephasing. Our approach utilizes the fact that environmental noise fluctuations are often slow compared to qubit dynamics, enabling us to solve for the qubit evolution analytically using the adiabatic theorem. Finding an analytical solution is crucial as it allows us to average these fluctuations with respect to any noise power spectrum by performing a Gaussian path integral. Consistency with the adiabatic approximation imposes constraints on the types of power spectra that may be treated in this way, which we derive below. We demonstrate our theory in the context of free induction decay (FID) in singlet-triplet (ST) qubits and obtain an analytical formula for the singlet return probability in the presence of both nuclear spin and charge noise. We compare our results with recent experiments reported in Ref.~\cite{Martins_PRL16}, finding excellent agreement. In addition, we show that a purely static noise model is also consistent with this data, illustrating the difficulty with conclusively ascertaining features of the bath from FID measurements. For alternative approaches to describing general types of classical noise, see Refs.~\cite{Green_NJP13,PazSilva_PRL14,Kabytayev_PRA14,Wang_PRB15}.

In developing our extension of the filter function formalism, we focus on the context of FID experiments with ST qubits in order to keep the discussion concrete, although the approach we develop has broad applicability and can be easily adapted to other systems or to include external control pulses. ST qubits \cite{Levy.02,Petta_Science05,Barthel_PRL09,Foletti_NP09,Bluhm_NP11,Maune_Nature12} are defined to live in the $S_z=0$ subspace of the two-spin Hilbert space of two separated electrons residing in a double quantum dot. A large external magnetic field is applied to energetically isolate this subspace from the states with $S_z = \pm 1$. Within the $S_z=0$ subspace, the qubit is subject to two effective fields: 1) tunneling through the inter-dot barrier gives rise to an exchange coupling $J$ between the two spins, and 2) a magnetic field gradient $h$, arising from either the polarization of lattice nuclear spins \cite{Foletti_NP09,Bluhm_PRL10,Barnes_PRL11,Rudner_PRB11,Economou_PRB14,Neder_PRB14} or a micromagnet \cite{Wu_PNAS14,Yoneda_APE15}, splits the energies of the states $\ket{\uparrow\downarrow}$ and $\ket{\downarrow\uparrow}$. Here the arrows indicate the spin projections of the electrons in the left and right dots, respectively, along the axis of the external magnetic field. The ST qubit Hamiltonian, expressed in the two-spin unpolarized triplet $\ket{T_0}$ and singlet $\ket{S}$ basis, is
\beq
H(t)=\frac{1}{2}\left(\begin{matrix} J+\delta J(t) & h+\delta h(t) \cr h+\delta h(t) & -J-\delta J(t) \end{matrix}\right).
\eeq
Here $\delta J(t)$ represents a small time-dependent fluctuation of the exchange coupling due to charge noise, while $\delta h(t)$ is a time-dependent fluctuation of the field gradient caused by hyperfine interactions with nuclear spins in the surrouding lattice. It is well established that nuclear spin (Overhauser) noise fluctuations are slow and must be treated as non-Markovian due to the long timescales of nuclear spin dynamics \cite{Yao_PRB06,Liu_NJP07,Cywinski_PRL09,Cywinski_PRB09,Coish_PRB10,Cywinski_APPA11,Barnes_PRL12b}. Although less is known about the origin of charge noise in quantum dots, there is evidence that it too has a non-Markovian character \cite{Dial_PRL13}. Moreover, it has been shown experimentally that both nuclear spin and charge noise in quantum dots are well described by classical Gaussian noise \cite{Medford_PRL12,Dial_PRL13}.

In the FID experiment of Ref.~\cite{Martins_PRL16}, the ST qubit is first initialized by tilting the double well potential into the (0,2) charge configuration and then waiting until the qubit relaxes into the singlet state.\cite{foot1} The double well potential is then adiabatically ramped back to a configuration favoring the (1, 1) charge state.  During the adiabatic ramp, the spin state evolves to $\ket{\uparrow\downarrow}$, the ground state for $|h| \gg J$ and $h<0$, the latter of which corresponds to a negative effective $g$-factor ($g=-0.44$ for GaAs) and a larger effective magnetic field in the left dot. At this point, the exchange coupling $J$ is rapidly increased by lowering the inter-dot barrier with a middle gate, and the qubit then evolves freely for a time $\tau$. After this free evolution, $J$ is quickly reduced and the adiabatic ramping is then reversed so that the $\ket{\uparrow\downarrow}$ component of the qubit state is mapped onto the (0,2) singlet while $\ket{\downarrow\uparrow}$ is mapped to the (1,1) triplet. The two states are thereby distinguished using a nearby charge sensor.

In this work, our goal is to calculate the probability that the state returns to $\ket{\uparrow\downarrow}$ after the free evolution time $\tau$, $P_{\uparrow\downarrow}(\tau)$, which is equal to the probability to be in the singlet state in the readout stage of the experiment provided that relaxation processes are negligible. We consider relaxation effects later on.

To incorporate the effects of noise into our calculation, we employ the filter function formalism. This formalism is developed by first solving for the evolution operator, using the result to compute observables such as $P_{\uparrow\downarrow}$ at arbitrary times, and then performing a Gaussian path integral over the noise fluctuations with an arbitrary noise power spectrum \cite{Cywinski_PRB08}.

Unlike in the case of pure dephasing, here we already run into difficulty at the first step: since the terms in $H$ corresponding to the exchange and gradient fields do not commute, it is impossible to solve exactly for the evolution for a $\delta J(t)$ and $\delta h(t)$ with arbitrary time-dependencies. To make any progress it is necessary to approximate the evolution; here we do so by invoking the slow timescales associated with both charge and nuclear spin noise, as compared with the those set by the fields $J$ and $h$.  We thereby work in the adiabatic limit, where the evolution generated by $H(t)$ can be approximated by computing the instantaneous eigenvalues and eigenvectors of $H(t)$ and combining them to obtain \cite{foot2}
\beq
U(t)\approx\left(\begin{matrix}\cos\phi-i\sin\phi\cos\chi & -i\sin\phi\sin\chi \cr -i\sin\phi\sin\chi & \cos\phi+i\sin\phi\cos\chi \end{matrix}\right),
\eeq
where
\bea
\chi(t)&\equiv&\arctan\left(\frac{h+\delta h(t)}{J+\delta J(t)}\right),\nn\\
\phi(t)&\equiv&\frac{1}{2}\int_0^tdt'[J+\delta J(t')]\sec(\chi(t')).
\eea
In terms of these functions, the adiabatic limit corresponds to $|\dot\chi|\ll2|\dot\phi|$. The probability to be in the $\ket{\uparrow\downarrow}$ state at the end of the free evolution is
\beq
P_{\uparrow\downarrow}(\tau)=1-\cos^2\chi(\tau)\sin^2\phi(\tau).\label{Pupdown}
\eeq

In order to facilitate the averaging over $\delta J$ and $\delta h$, we use the assumption that these fluctuations are small and expand $\chi$ and $\phi$ to leading order:
\bea
\chi&\approx& \bar\chi+\cos^2\bar\chi\frac{\delta h}{J}-\sin(2\bar\chi)\frac{\delta J}{2J},\nn\\
\dot\phi&\approx& \frac{J}{2}\sec\bar\chi+\frac{1}{2}\cos\bar\chi\delta J+\frac{1}{2}\sin\bar\chi\delta h,\label{expansion}
\eea
where $\bar\chi\equiv\arctan(h/J)$. Using these expansions and taking averages over the noise by performing Gaussian path integrals, we find
\bea
&&\!\!\!\!\!\!\langle P_{\uparrow\downarrow}(\tau)\rangle\approx1-\tfrac{1}{2}\cos^2\bar\chi[1-\cos(J\tau\sec\bar\chi)c_J(\tau)c_h(\tau)]\nn\\&& \!\!\!\!\!\!-\tfrac{\cos\bar\chi}{J}\sin(J\tau\sec\bar\chi)[\cos^2\bar\chi c_J(\tau)\dot c_h(\tau)-\sin^2\bar\chi \dot c_J(\tau) c_h(\tau)],\nn\\
&&c_J(\tau)\equiv\exp\left\{-2\cos^2\bar\chi\int_0^\infty \frac{d\omega}{\pi}S_J(\omega)\frac{\sin^2(\omega \tau/2)}{\omega^2}\right\},\nn\\
&&c_h(\tau)\equiv\exp\left\{-2\sin^2\bar\chi\int_0^\infty \frac{d\omega}{\pi}S_h(\omega)\frac{\sin^2(\omega \tau/2)}{\omega^2}\right\},\label{Pupdownave}
\eea
where $S_J$ and $S_h$ are the power spectra for charge and nuclear spin noise, respectively. Here, we are neglecting potential cross-correlations between the two noise sources for simplicity. For small fluctuations, the terms in the first line of Eq.~\eqref{Pupdownave} dominate, and the probability evolves as though there were a single noise source with an effective power spectrum $S_{\rm eff}(\omega)=\cos^2\bar\chi S_J(\omega)+\sin^2\bar\chi S_h(\omega)$, where the relative importance of the different noise contributions is weighted by the ratio $h/J$ through the dependence on $\bar\chi$. The second line of Eq.~\eqref{Pupdownave} incorporates additional small transient effects at short times. Although the path integrals were performed without making assumptions about $S_J(\omega)$ and $S_h(\omega)$, some restrictions must be imposed on these functions to ensure the validity of the adiabatic and small-fluctuation approximations as we discuss in detail below.

As a test of our main result, Eq.~\eqref{Pupdownave}, we first compare its predictions against an exact solution, which is possible in the quasistatic limit where both the charge and nuclear spin noise are treated as random, time-independent offset fields. To obtain the exact solution for the evolution with quasistatic noise we set $\delta J=\delta h=0$ in Eq.~\eqref{Pupdown}, giving $P_{\uparrow\downarrow}(\tau)=1-\tfrac{1}{2}\cos^2\chi\left[1-\cos(J\tau\sec\chi)\right]$. Quasistatic Gaussian noise is then treated by taking Gaussian averages over $h$ and $J$ with standard deviations $\sigma_h$ and $\sigma_J$. Using $\chi$ rather than $h$ as an integration variable enables us to perform the $J$ integration exactly, yielding $\langle P_{\uparrow\downarrow}(\tau)\rangle=M(\tau)+P_\infty$ where
\beq
P_\infty=1-\frac{e^{-\tfrac{h_0^2}{2\sigma_h^2}-\tfrac{J_0^2}{2\sigma_J^2}}}{\sqrt{\pi}\sigma_h\sigma_J}\int_{-\pi/2}^{\pi/2} d\chi\frac{b(\chi)}{[a(\chi)]^{3/2}}\exp\left(\frac{b(\chi)^2}{a(\chi)}\right),\label{Pinf}
\eeq
\bea
&&\!\!\!\!\!\!M(\tau)=\frac{e^{-\tfrac{h_0^2}{2\sigma_h^2}-\tfrac{J_0^2}{2\sigma_J^2}}}{\sqrt{\pi}\sigma_h\sigma_J} \nn\\&&\!\!\!\!\!\!\!\!\times\int_{-\pi/2}^{\pi/2} d\chi\hbox{Re}\left[\frac{b(\chi){+}i\tau\sec\chi}{[a(\chi)]^{3/2}}\exp\left(\frac{[b(\chi){+}i\tau\sec\chi]^2}{a(\chi)}\right)\right],\nn\\&&\label{Moftau}
\eea
with $a(\chi){=}2\tan^2\chi/\sigma_h^2{+}2/\sigma_J^2$,$b(\chi){=}h_0\tan\chi/\sigma_h^2{+}J_0/\sigma_J^2$.

\begin{figure}
\includegraphics[width=\columnwidth]{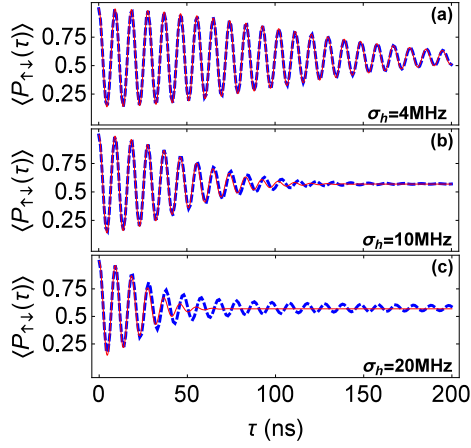}
\caption{\label{fig:FID1} Quasistatic limit of Eq.~\eqref{Pupdownave} (red) vs. exact quasistatic model from Eqs.~\eqref{Pinf}, \eqref{Moftau} (blue, dashed) for several values of $\sigma_h$ with $J=100$MHz, $\sigma_J=100$kHz, $h_0=40$MHz.}
\end{figure}

Returning to the approximate evolution described by Eq.~\eqref{Pupdownave}, the quasistatic limit corresponds to taking a noise power spectrum $S(\omega)=2\pi\sigma^2\delta(\omega)$, where $\sigma$ is the width of the Gaussian distribution. When both nuclear and charge noise are quasistatic, we have $c_J(\tau)=e^{-\cos^2\bar\chi\sigma_J^2\tau^2/2}$ and $c_h(\tau)=e^{-\sin^2\bar\chi\sigma_h^2\tau^2/2}$, implying an effective dephasing time $T_2^*=\sqrt{2}/\sqrt{\cos^2\bar\chi\sigma_J^2+\sin^2\bar\chi\sigma_h^2}$. A comparison of the resulting $\langle P_{\uparrow\downarrow}(\tau)\rangle$ from Eq.~\eqref{Pupdownave} with the exact quasistatic evolution, Eqs.~\eqref{Pinf} and \eqref{Moftau}, is shown in Fig.~\ref{fig:FID1}. We see that good agreement between the two occurs when $\sigma_h\ll |h_0|$, verifying the consistency of our results in the regime of small fluctuations.

\begin{figure}
\includegraphics[width=\columnwidth]{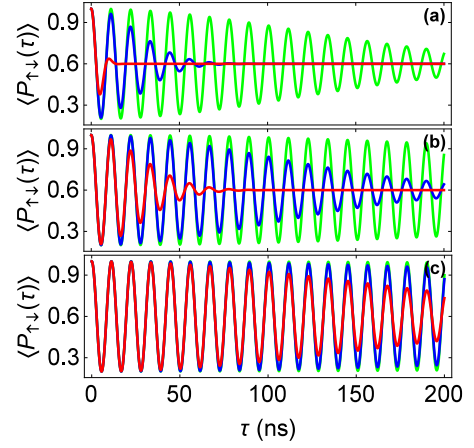}
\caption{\label{fig:FID2} FID for ST qubit subject to $1/\omega^{\alpha}$ noise with $J_0=80$MHz, $h_0=40$MHz, $\omega_{ir}=10$kHz, $\omega_{uv}=100$kHz for (a) $A=400$kHz, (b) $A=200$kHz, (c) $A=100$kHz and with $\alpha=2$ (green), $\alpha=3$ (blue), $\alpha=4$ (red).}
\end{figure}

Both nuclear spin and charge noise in quantum dots have been shown to be well described by power-law spectra, $S(\omega)=A^{1+\alpha}/\omega^\alpha$. In this case, the frequency integration in Eq.~\eqref{Pupdownave} can be performed exactly:
\bea
&&\!\!\!\!\!\!G(\tau,\alpha,\omega_{ir},\omega_{uv})\equiv\frac{2}{\pi}\int_{\omega_{ir}}^{\omega_{uv}}d\omega\frac{\sin^2(\omega \tau/2)}{\omega^{2+\alpha}}\\&&\!\!\!\!\!{=}\frac{\hbox{Re}[E_{\alpha
   +2}(i \tau \omega_{uv})]}{\pi\omega_{uv}^{\alpha+1}}{-}
   \frac{\hbox{Re}[E_{\alpha +2}(i \tau \omega_{ir})]}{\pi\omega_{ir}^{\alpha+1}}{+}\frac{\omega_{ir}^{-\alpha -1}{-}\omega_{uv}^{-\alpha -1}}{\pi(\alpha+1)},\nn
\eea
where $E_\beta(z)=\int_1^\infty duu^{-\beta} e^{-zu}$ is the exponential integral function, and $\omega_{ir}$, $\omega_{uv}$ are infrared and ultraviolet cutoffs of the noise spectrum. The resulting return probabilities for a single effective power spectrum $S_{\rm eff}(\omega)$ for different values of $\alpha$ and $A$ are shown in Fig.~\ref{fig:FID2}.

In order to determine the relevance of Eq.~\eqref{Pupdownave} for real experiments, we must return to the question of how the power spectrum is constrained by the adiabatic and small-fluctuation approximations. First we consider the adiabaticity condition, $|\dot\chi|\ll2|\dot\phi|$.  In terms of Hamiltonian parameters, this condition is $|h||\delta \dot J(t)|,J|\delta\dot h(t)|\ll(h^2+J^2)^{3/2}$. We interpret this as the condition that the noise-averaged quantities, $|h|\langle |\delta \dot J(t)|\rangle$ and $J\langle|\delta\dot h(t)|\rangle$, are small compared to $(J^2+h^2)^{3/2}$. These quantities can be computed exactly: $\langle |\delta \dot J(t)|\rangle=\frac{1}{\pi}\sqrt{\int_{-\infty}^\infty d\omega \omega^2S_J(\omega)}$, and similarly for $\langle |\delta \dot h(t)|\rangle$ with $S_J$ replaced by $S_h$. Adiabaticity thus imposes the following constraints on the noise spectra:
\beq
\sqrt{\int_{-\infty}^\infty d\omega \omega^2S_J(\omega)}\ll\frac{\pi(h^2+J^2)^{3/2}}{|h|},
\eeq
and similarly for $S_h(\omega)$ with $h\leftrightarrow J$ on the right-hand side. In the case of a power-law spectrum, this becomes
\beq
A^{\tfrac{1+\alpha}{2}}\sqrt{\frac{\omega_{uv}^{3-\alpha}-\omega_{ir}^{3-\alpha}}{3-\alpha}}\ll\frac{\pi(h^2+J^2)^{3/2}}{|h|}.\label{adiabconstraint}
\eeq
For the parameters of Ref.~\cite{Medford_PRL12}, where it was found that for nuclear spin noise $\alpha=2.6$, $A\approx0.3$MHz, $\omega_{ir}\sim10$kHz, $\omega_{uv}\sim100$kHz, the left-hand side is $\sim90$kHz$^2$, easily satisfying the constraint for typical $h$ and $J$ in the MHz to GHz range. Ref.~\cite{Dial_PRL13} measured a charge noise spectrum with $\alpha=0.7$, $A\sim100$MHz, $\omega_{ir}\sim50$kHz, $\omega_{uv}\sim1$MHz, giving $\sim30$MHz$^2$ for the left-hand side of Eq.~\eqref{adiabconstraint}, again a fairly weak constraint.

A second constraint on the noise spectra comes from requiring the fluctuation terms in Eq.~\eqref{expansion} to be small compared to the leading order: $J|\delta J(t)|$, $|h||\delta h(t)|\ll h^2+J^2$. We again interpret these as constraints on the noise-averaged quantities, given exactly by $\langle |\delta J(t)|\rangle=\frac{1}{\pi}\sqrt{\int_{-\infty}^\infty d\omega S_J(\omega)}$, and similarly for $\langle |\delta h(t)|\rangle$ with $S_J\leftrightarrow S_h$. The small-fluctuation approximation thus requires
\beq
\sqrt{\int_{-\infty}^\infty d\omega S_J(\omega)}\ll \frac{\pi(h^2+J^2)}{J},
\eeq
and similarly for $S_h$ with $h\leftrightarrow J$. In the case of a power-law spectrum, this becomes
\beq
A^{\tfrac{1{+}\alpha}{2}}\sqrt{\frac{\omega_{ir}^{1{-}\alpha}{-}\omega_{uv}^{1{-}\alpha}}{\alpha-1}}\ll \frac{\pi(h^2+J^2)}{J},\frac{\pi(h^2+J^2)}{|h|}.
\eeq
For the parameters of Ref.~\cite{Medford_PRL12}, the left-hand side evaluates to $\sim3$MHz, while those from Ref.~\cite{Dial_PRL13} give $\sim70$MHz. Although the latter does restrict $J$ somewhat, the small-fluctuation assumption still holds for experimentally relevant $J$'s on the order of hundreds of MHz to GHz.

In the case of nuclear spin noise, the validity of the small-fluctuation approximation can depend on whether the nuclear spins are in a thermal or narrowed polarization state. If the distribution of nuclear polarization is sufficiently narrow (in the quasistatic limit, this means $\sigma_h\ll |h_0|$), then the small-fluctuation approximation should be valid. However, for a thermal state, the width of the distribution is comparable to the mean polarization as in Ref.~\cite{Martins_PRL16}, in which case this approximation becomes questionable. In this situation, it is still possible to obtain analytical results for the return probability at times less than the timescale of nuclear spin dynamics (typically microseconds) by treating the nuclear noise as quasistatic. In this case, we have
\bea
&&\!\!\!\!\!\!\langle P_{\uparrow\downarrow}(\tau)\rangle = 1-\frac{J}{2\sqrt{2\pi}\sigma_h}\int_{-\pi/2}^{\pi/2}d\bar\chi e^{-\frac{(J\tan\bar\chi-h_0)^2}{2\sigma_h^2}}\nn\\&&\!\!\!\!\!\!\times\bigg[1{-}\cos\left(J\tau\sec\bar\chi\right)\exp\left\{{-}A^{1{+}\alpha} \cos^2\bar\chi G(\tau,\alpha,\omega_{ir},\omega_{uv})\right\}\bigg].\nn\\&&\label{Pupdownquasi}
\eea
Eq.~\eqref{Pupdownquasi} gives the return probability for a ST qubit subject to quasistatic nuclear noise and $1/\omega^\alpha$ charge noise.

\begin{figure}
\includegraphics[width=\columnwidth]{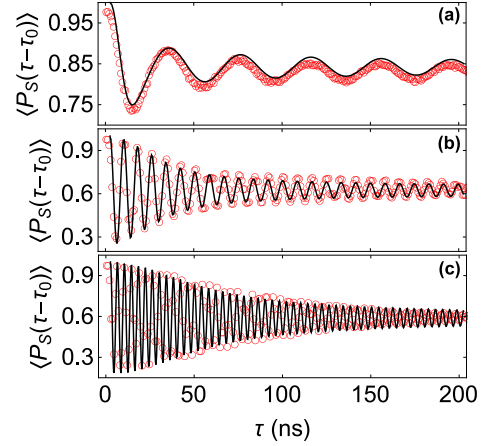}
\caption{\label{fig:FID3} FID for ST qubit subject to $1/f^{\alpha}$ charge noise and quasistatic nuclear noise with $\alpha=2.5$, $\omega_{ir}=5$Hz, $\omega_{uv}=\infty$, $h_0=40$MHz, $\sigma_h=23$MHz, $T_1=30\mu$s, $\tau_0=$2.6ns, and $A=2\times10^{-5}J$ for (a) $J=25.2$MHz, (b) $J=121$MHz, (c) $J=249$MHz. Experimental data from Ref.~\cite{Martins_PRL16} shown in red, simulation in black.}
\end{figure}

In order to quantitatively compare this result to the experiment of Ref.~\cite{Martins_PRL16}, we must account for the effects of spin relaxation during readout. For a relaxation time $T_1$ and measurement time $T_M$, the measured singlet return probability becomes $P_S(\tau)=1-\frac{T_1}{T_M}(1{-}e^{-T_M/T_1})[1-P_{\uparrow\downarrow}(\tau)]$ \cite{Reilly_PRL08}. Combining this with Eq.~\eqref{Pupdownquasi}, we set $T_M=10\mu$s and choose $A$ to depend linearly on $J$ to reflect the fact that $J$ is more sensitive to noise when the voltage on the middle gate is larger, corresponding to larger $J$. In Fig.~\ref{fig:FID3}, we fit the parameters $\alpha$, $A/J$, $\omega_{ir}$ and include an additional phase shift, $\tau_0$, to account for a finite rise time in the pulse generation. We find good agreement with the experiment of Ref.~\cite{Martins_PRL16} using measured values of $h_0$, $\sigma_h$, and $T_1$. Interestingly, in the experimental parameter regime, the frequency of oscillations is given by $J$ and does not depend on the Overhauser field or other parameters. This is most easily understood by considering an infinitely broad Overhauser distribution in the limit of weak charge noise, in which case the oscillatory part of Eq.~\eqref{Pupdownquasi} is $\int_{-\pi/4}^{\pi/4} d\bar\chi\cos\left(J\tau\sec\bar\chi\right)\approx\sqrt{\tfrac{2\pi}{J\tau}}\cos(J\tau+\pi/4)$. The fact that the oscillation frequency is given by $J$ makes it straightforward to extract $J$ experimentally.

For completeness, we point out that good agreement with the experimental data can also be achieved using a purely quasistatic model for both nuclear spin and charge noise.  For the same values of $h_0$, $\sigma_h$ and $J$ used in Fig.~\ref{fig:FID3}, we find that the quasistatic theory, Eqs.~\eqref{Pinf}, \eqref{Moftau}, agrees closely with the experiment for $\sigma_J=4.26\times10^{-3}J$. The fact that a purely quasistatic model is also consistent with the data highlights the difficulty with discerning properties of the noise from FID experiments.

In conclusion, we have developed an analytical approach to determine the free evolution of a qubit exposed to multiple, simultaneous noise sources with non-trivial power spectra. Our method works in situations where the system is subject to small, non-Markovian noise fluctuations, and we have derived constraints on the noise spectra to determine when these conditions are fulfilled. We have shown that larger noise fluctuations can be included by employing the quasistatic bath approximation, and we found that our resulting theory agrees well with recent experiments. In future work, we will extend our approach to include external driving in order to support experimental efforts to better understand the nature and consequences of decoherence in qubit systems.

\acknowledgements
This work was supported by IARPA-MQCO, LPS-MPO-CMTC, the Villum Foundation, the EC FP7- ICT project SiSPIN no. 323841, the Army Research Office and the Danish National Research Foundation.


\begin{thebibliography}{40}
\bibitem{foot1}We assume any initial correlations between electronic and nuclear spins are negligible.
\bibitem{foot2}Here we ignore geometric phase contributions, which vanish at linear order in the noise amplitudes. See Ref.~\cite{SanJose_PhysE07}.
\end{thebibliography}
\end{document}